\documentclass[aps,prl,twocolumn,showpacs]{revtex4}

\usepackage{epsfig}

\newcommand{\lbfig}[1]{\refstepcounter{fig} \label{#1} }
\newcounter{fig}

\newcommand{\nc}{\newcommand}
\nc{\be}{\begin{equation}}
\nc{\ee}{\end{equation}}
\nc{\bea}{\begin{eqnarray}}
\nc{\eea}{\end{eqnarray}}
\nc{\bi}[1]{\bibitem{#1}}
\nc{\lsim}{\mbox{\raisebox{-.6ex}{~$\stackrel{<}{\sim}$~}}}
\nc{\gsim}{\mbox{\raisebox{-.6ex}{~$\stackrel{>}{\sim}$~}}}

\begin{document}

%\rightline{HD-THEP-02-14}
%\rightline{UFIFT-HEP-02-03}
%\rightline{\today}

\leftline{\hspace{3.4in} HD-THEP-02-14, UFIFT-HEP-02-03, Imperial/TP/1-02/21}

\vskip 0.2in

\title{Photon mass from inflation}

\author{Tomislav Prokopec}
\email[]{T.Prokopec@thphys.uni-heidelberg.de}
\affiliation{Institut f\"ur Theoretische Physik, Heidelberg University,
             Philosophenweg 16, D-69120 Heidelberg, Germany}

\author{Ola T\"{o}rnkvist}
\email[]{o.tornkvist@ic.ac.uk}
\affiliation{Theoretical Physics Group, Imperial College,
             Prince Consort Road, London SW7 2BZ, UK}

\author{Richard Woodard}
\email[]{woodard@phys.ufl.edu}
\affiliation{Department of Physics, University of Florida,
             Gainesville, FL 32611, USA}

\date{\today}

\begin{abstract}

 We consider vacuum polarization from massless scalar electrodynamics 
in de Sitter inflation. The theory exhibits a 3+1 dimensional analogue of the 
Schwinger mechanism in which a photon mass is dynamically generated.  The
mechanism is generic
for light scalar fields that couple minimally to gravity.
The non-vanishing of the photon mass 
during inflation may result in magnetic fields
on cosmological scales.

\end{abstract}

\pacs{98.80.Cq, 98.80.Hw, 04.62.+v}

\maketitle

%
%%%%%%%%%%%%%%%%%%%%%%%%%%%%%%%%%%%%%%%%%%%%%%%%%%%%%%%%%%%%%%%%%%%%%%%%%%%%%%%
%  MAIN TEXT
%%%%%%%%%%%%%%%%%%%%%%%%%%%%%%%%%%%%%%%%%%%%%%%%%%%%%%%%%%%%%%%%%%%%%%%%%%%%%%%
%

%\section{Introduction}

{\it 1. Introduction.}
The mass of the photon has been under scrutiny from the early days of
quantum mechanics~\cite{deBroglie:1940-Schrodinger:1943-BassSchrodinger:1955},
and this has resulted in stringent limits. The best laboratory bounds of 
$m_\gamma \lsim 10^{-46}{\rm g}\approx 10^{-14}$~eV
are derived from measurements of potential deviations from the Coulomb 
law~\cite{WilliamsFallerHill:1971}. 
The most precise direct bounds are based on measurements
of Earth's magnetic field~\cite{FischbachKloorLangelLiuPeredo:1994}
and the Pioneer-10 measurements of Jupiter's magnetic 
field~\cite{DavisGoldhaberNieto:1975} and yield 
$m_\gamma \lsim 10^{-15}$~eV. For a review
of other methods and limits, see~\cite{GoldhaberNieto:1971}.

Although there is little direct evidence about the photon mass before
the time of 
matter-radiation
decoupling, it is usually assumed to 
%be 
have been
equally small on
the basis of 
%the 
current (approximately flat space) data, the conformal
invariance of classical electromagnetism, and the 
%fact 
deduction
that the geometry
of the early universe was conformally flat to a high degree. 
%However, 
It is
 well known,
how\-ever,
that quantum electrodynamic (QED) corrections break conformal
invariance in curved space~\cite{DrummondHathrell,Dolgov:1981}.
This may induce important effects in the early
universe~\cite{Prokopec:2001,TurnerWidrow:1988,Dolgov:1993}. 

The problem of full nonlocal vacuum polarization induced by 
matter loops in curved spacetimes has so far not been considered. 
That is precisely the subject of this work. We show that, 
in a locally-de-Sitter inflationary spacetime and
in the presence of a light, minimally coupled, charged scalar field,
%the vacuum gets polarized and consequently the photon acquires 
%a mass at the one-loop level. 
the polarization of the vacuum induces a photon mass at the 
one-loop level.
The effect is caused by the coup\-ling of the gauge
field to 
infrared scalar 
modes
%fields 
that generically 
undergo superadiabatic amplification on superhorizon scales.
%on superhorizon scales by superadiabatic amplification. 
This 
represents 
%provides 
a novel mechanism by
which gauge fields can become massive; 
it 
is analogous to the 
Schwinger mechanism~\cite{Schwinger:1962}, according to which the photon 
of 1+1 dimensional QED acquires a mass $m_\gamma = e/\sqrt{\pi}$.
%The mechanism we consider here is a 3+1 dimensional 
%analog of the Schwinger mechanism. 
The photon vacuum polarization 
%we 
calculated here 
incorporates this and other known effects \cite{Prokopec:2001,Dolgov:1993}
%contains as special cases the above mentioned
%effects 
%{\bf Check??}
on the photon dynamics in scalar QED.

 The cosmological relevance of our findings stems from recent
work~\cite{DimopoulosProkopecTornkvistDavis:2001,
DavisDimopoulosProkopecTornkvist:2000}, where it was argued that a 
dynamically generated gauge-field mass in inflation may result in
the
generation of large-scale magnetic fields, which could seed the galactic 
dynamo and thus offer an explanation for the micro-Gauss-strength 
galactic magnetic fields observed today~\cite{Kronberg:1994}. 
An analogous effect arises in a more conventional Higgs mechanism 
realised in inflation~\cite{TornkvistDavisDimopoulosProkopec:2000}.
The resulting magnetic field spectrum is 
%generically 
of the form
$B_{\ell}\propto \ell^{-1}$, where $\ell$ is 
the correlation length. 
%which 
This can be sufficiently strong 
to seed the galactic dynamo mechanism~\cite{dynamo}
in flat universes with a dark-energy 
component~\cite{DavisLilleyTornkvist}. 
For reviews of other mechanisms that may generate large-scale magnetic 
fields, see~\cite{Prokopec:2001,GrassoRubinstein:2000,TurnerWidrow:1988}.

The authors of Refs.~\cite{DavisDimopoulosProkopecTornkvist:2000,
DimopoulosProkopecTornkvistDavis:2001} have used a 
mean-field approximation
to model the backreaction of superhorizon scalar fields on gauge fields. 
Their analysis indicates that the photon acquires a mass in inflation. 
In this Letter, we calculate the gauge-invariant photon self-energy 
at the one-loop level, from which we obtain the photon mass. 
Our perturbative result 
is 
%found to be 
%identical to
in 
%rough 
agreement with 
%corroborates the result of
that of
the mean-field analysis
in~\cite{DimopoulosProkopecTornkvistDavis:2001,
DavisDimopoulosProkopecTornkvist:2000}.

% \section{Scalar electrodynamics in de Sitter inflation}

\vskip 0.1in

{\it 2. The model.} 
%In this Letter we c
Consider scalar electrodynamics
in de Sitter inflation with the Lagrangean
\begin{eqnarray}
\!\!{\cal L}_{\rm \phi ED}
 \! =\! -\frac 14 \sqrt{-g}g^{\mu\rho}g^{\nu\sigma}
           F_{\mu\nu}F_{\rho\sigma}
    \!-\! \sqrt{-g}g^{\mu\nu}(D_\mu\phi)^\dagger D_\nu\phi,
%\nonumber\\ 
\label{ph-m.2}
\end{eqnarray}
where $D_\mu=\partial_\mu+ieA_\mu$ is the covariant derivative,
$F_{\mu\nu}=\partial_\mu A_\nu-\partial_\nu A_\mu$ is the gauge field strength,
$g_{\mu\nu}=a^{2}\eta_{\mu\nu}$ is a conformally flat metric 
and $g = {\rm det}[g_{\mu\nu}] = 
%-a^8$.
-a^{2D}$.
%The actual 
Our calculation was 
%done
performed
in $D$ spacetime dimensions
using dimensional regularization \cite{ProkopecTornkvistWoodard:2002}.
However, with only two minor modifications it can be understood by
working in $D=4$.

We require that the scalar field be light
in comparison to the Hubble parameter $H$, $m_\phi\ll H\sim 10^{13}$~GeV,
so that scalar-field perturbations may grow during inflation.
The current experimental bounds on the mass of a charged scalar particle
$m_\phi \gsim M_{\rm EW}\sim 10^{2}~{\rm GeV}$ can be amply satisfied.
The obvious candidates for $\phi$ are the charged Higgs particles and 
the supersymmetric partners of the Standard-Model leptons and quarks.

%In $D=4$ t
The %Feynman 
scalar propagator 
%is 
$i\Delta (x,x')\! =\!
\langle\Omega| T [\phi^\dagger(x)\phi(x')] |\Omega\rangle$,
where $|\Omega\rangle$ denotes the Bunch-Davies vacuum,
%In $D=4$ it obeys 
satisfies for $D=4$
the equation
%satisfies in conformally flat spacetimes the 
%equation (with $m_\phi=0$)
%
\begin{equation}
\partial^{\mu} \Big( a^2\partial_\mu i\Delta (x,x') \Big) = i\delta^4(x-x'),
\label{ph-m.3}
\end{equation}
where the raising of indices is from now on defined as  
$\partial^{\mu}\equiv \eta^{\mu\nu}\partial_\mu$.
%This convention is suitable for spacetimes that respect conformal 
%symmetry. We assumed that $\phi$ couples minimally to gravity. 
In the de Sitter spacetime, where the scale factor is given by 
$a = -1/H\eta$, $H$ denotes the Hubble parameter and $\eta$ denotes
conformal time, one can show that the solution of~(\ref{ph-m.3}) 
reads~\cite{FordParker:1977+}
\begin{equation}
i\Delta (x,x') = \frac{H^2}{4\pi^2}\left(
    \frac{\eta\eta'}{\Delta x^2} - \frac{1}{2}\ln (H^2\Delta x^2) \right),
\label{ph-m.4}
\end{equation}
where 
%$\Delta x^2 \equiv - (\eta-\eta')^2 + \vec x-{\vec x}\,'^2  + i\epsilon$
$\Delta x^2 \equiv - (|\eta-\eta'| - i\epsilon)^2 
                 + \vert\vec x-{\vec x}\,'\vert^2 $.
Our metric convention is $\eta_{\mu\nu} = {\rm diag}[-1,1,1,1]$
and $x^\mu = (x^0,\vec x)$, $x^0\equiv \eta$,
$\partial_\mu = (\partial_0,\partial_x,\partial_y,\partial_z)$.

 On the other hand, 
the propagation of 
free photons 
%the photon field 
in de Sitter inflation is governed,
on the classical level,
by the flat-space Maxwell equations,
%in the Minkowski vacuum, 
$\partial^\nu F_{\nu\mu}  = 0$.

%\section{Photon self-energy}

\vskip 0.1in

{\it 3. Photon self-energy.} Consider now the photon self-energy,
which 
acquires one-loop level contributions  
from the diagrams shown in Fig.~\ref{figure 1} and 
can be written as
\begin{eqnarray}
\lefteqn{\!i[^\mu\Pi^{\nu}](x,x') = 
  -2i e^2 aa' i\Delta(x,x')\, \eta^{\mu\nu} \delta^4(x-x')}
\quad\quad~
\nonumber\\
&+& 2e^2 \eta^{\mu\rho}\eta^{\nu\sigma}\!
\Big(\!
  \partial_\rho [aa'i\Delta(x',x)]
           \partial'_\sigma[aa'i\Delta(x,x')]
\nonumber\\
&-& [aa'i\Delta(x',x)] \partial_\rho \partial'_\sigma[aa'i\Delta(x,x')]
\Big),\quad\;
\label{ph-m.8}
\end{eqnarray}
where $\partial_\rho \equiv\partial/\partial x^\rho$,
$\partial'_\sigma \equiv\partial/\partial {x'}^\sigma$,
$a=a(\eta)$, $a'=a(\eta')$, we used the symmetry
$i\Delta(x,x') = i\Delta(x',x)$ of~(\ref{ph-m.4}) and neglected
for the moment the contribution from the counterterm in Fig.~\ref{figure 1}.
%The scale factors in~(\ref{ph-m.8}) are placed inside the arguments of the 
%derivatives so that conformal invariance is manifest. 
%
\begin{figure}[htbp]
\begin{center}
\epsfig{file=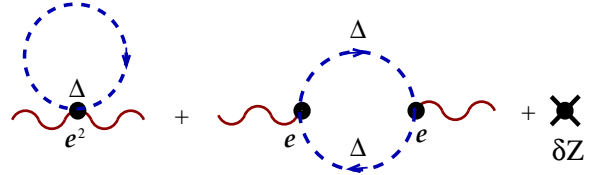, height=0.9in,width=3.in}
\end{center}
\vskip -0.15in
\lbfig{figure 1}
\caption[fig1]{%
\small
The one-loop diagrams contributing to the photon self-energy.
}
\end{figure}

After some algebra, the one-loop self-energy  
$[^\mu\Pi^{\nu}]$ can be recast in the form 
\begin{eqnarray}
i[^\mu\Pi^{\nu}](x,x') &=& 
  \eta^{\mu\rho}\eta^{\nu\sigma}
 \Big(
    \eta_{\rho\sigma}\partial'\cdot\partial - \partial'_\rho\partial_\sigma
 \Big) i\Pi^{(1)}(x,x')
\nonumber\\*
\!\!\!\!\! &+& \!\!
  \eta^{\mu i}\eta^{\nu j}
 \Big(
    \eta_{ij}\partial_l'\partial_l - \partial_i'\partial_j
 \Big) i\Pi^{(2)}(x,x'),\!\!
\label{ph-m.9}
\end{eqnarray}
where 
\begin{eqnarray}
\!\!\!  i\Pi^{(1)}(x,x') \!\!\!&=&\!\!\! \frac{e^2}{8\pi^4}
 \!\left[
   \frac{1}{6\Delta x^4} 
 \!-\! \frac{1}{\eta\eta'}
 \left(\!
  \frac{1}{\Delta x^2}\!+\! \frac{\ln(H^2\Delta x^2)}{2\Delta x^2}
 \!\right)
 \right]
%\nonumber\\
\label{ph-m.10}
\\*
  i\Pi^{(2)}(x,x') \!\!\!&=&\!\!\! \frac{e^2}{8\pi^4}
 \frac{1}{\eta^2{\eta'}^2}
   \! \left[\frac{1}{2}\ln(H^2\Delta x^2)
            \!+\! \frac{1}{8}\ln^2(H^2\Delta x^2)\right].
\nonumber\\
\label{ph-m.11}
\end{eqnarray}
The first term in~(\ref{ph-m.10}) is the standard Minkowski
vacuum contribution. This term is singular in the (ultraviolet) coincident
limit $x \rightarrow x'$, while the other terms originate from 
the
nonconformal
coupling of scalar fields to gravity in 
the
de Sitter background and 
are completely integrable. 
The ultraviolet problems are resolved 
by using dimensional regularization, that is by calculating 
in $D$ spacetime dimensions and, subsequently, by renormalizing the
self-energy.
The result of this rather technical
analysis, which we present in~\cite{ProkopecTornkvistWoodard:2002}, is 
that $i\Pi^{(1)}\to i \Pi^{(1)}_{\rm ren}$~, where
\begin{eqnarray}
&& \!\!\!\! i\Pi^{(1)}_{\rm ren}(x,x') = 
i\,\delta \Pi_{\rm anom}(x,x')
+
\frac{e^2}{8\pi^4}
 \bigg[
   \partial'\cdot\partial\,\frac{\ln (\mu^2\Delta x^2)}{24\Delta x^2} 
\nonumber\\
 &&+\; \frac{1}{\eta\eta'}\partial'\cdot\partial
    \left(\frac{1}{16}\ln^2(H^2\Delta x^2) 
 + \frac 18\ln(H^2\Delta x^2)\right)
 \bigg],\quad 
\label{ph-m.12}
\end{eqnarray}
%
%where
and 
$i\Pi^{(2)}_{\rm ren} = i\Pi^{(2)}$ remains unchanged.
Here, $\mu$ is the renormalization scale and 
%$
\begin{equation}
i\,\delta \Pi_{\rm anom}(x,x')
=-i\,\frac{\alpha_e}{6\pi}\,\ln(a)\, \delta^4(x-x')
\end{equation}
%$
[with $\alpha_e=e^2/4\pi$] 
is a local, anomalous contribution resulting from an imperfect
cancellation in expanding backgrounds between the local term and
counterterm in Fig.~1.

%
%Furthermore, it can be easily shown with dimensional regularization
%that, as a consequence of an imperfect cancellation 
%in expanding backgrounds
%between the local 
%term
%and counterterm in Fig.~\ref{figure 1},
% in expanding backgrounds, 
%there is a local anomalous contribution to the vacuum polarization 
%of the form
%%
%\begin{eqnarray}
%  i\delta [^\mu\Pi^\nu]_{\rm anom}(x,x') &=& -i\frac{\alpha_e}{6\pi}
% \Big(
%    \eta^{\mu\nu} \partial'\cdot\partial - {\partial^\mu}'\partial^\nu
% \Big)
%\nonumber\\
%&\times& \Big(\ln (a)\delta^4(x-x')\Big),
%\label{ph-m.12a}
%\end{eqnarray}
%
%where $\alpha_e = e^2/4\pi$.
Upon combining the classical action $S_0$ with the anomaly contribution
$\delta S_{\rm anom}$, we get
\begin{eqnarray}
\!\!\!&&\!\!\!\!\!\!\!\! S_0 + \delta S_{\rm anom} = 
 - \frac 12\int d^4 x d^4 x'
 A_\mu(x) 
 \bigg\{(\eta^{\mu\nu}\partial'\cdot\, \partial - {\partial^\mu}'\partial^\nu) 
\nonumber\\
 &&\quad\quad\times
   \bigg[\bigg(1+\frac{\alpha_e}{6\pi}\ln\Big(\frac{a}{a_0}\Big)\bigg) 
      \delta^4(x-x')\bigg]\bigg\}
 A_\nu(x') .
\label{anomaly-action}
\end{eqnarray}
This is the scalar electrodynamics equivalent of the Dolgov
anomaly~\cite{Dolgov:1981,MazzitelliSpedalieri:1995}. 
Since the anomalous contribution  to the effective action is proportional
to $\ln a$, we 
infer 
%expect
that the anomaly affects the photon dynamics quite 
mildly~\cite{Prokopec:2001,MazzitelliSpedalieri:1995} when compared with 
the effect of 
the photon mass, which contributes as $\propto a^2$
and hence is parametrically much larger. 

The transverse structure of the vacuum polarization~(\ref{ph-m.9})
implies that the Ward identities 
$\partial_\mu [^\mu\Pi^{\nu}] = 0 = \partial'_\nu [^\mu\Pi^{\nu}]$
%
%\begin{equation}
%       \partial_\mu [^\mu\Pi^{\nu}] = \partial'_\nu [^\mu\Pi^{\nu}] = 0
%\label{ph-m.13}
%\end{equation}
%
are obeyed, so that gauge invariance remains unbroken. The structure
of our result~(\ref{ph-m.9})-(\ref{ph-m.12}) is very similar to that of
thermal QED~\cite{LeBellac:1996}, which may have something to do with
regarding inflationary particle production as Hawking radiation.
The spacetime generalization of the standard thermal
transverse and `longitudinal' projectors are 
\begin{eqnarray} 
 \!\! P_{\rm T}^{\mu\nu} 
  \!\!=\! \eta^{\mu i}\eta^{\nu i}\!
    \left(\delta_{ij} \!-\! \frac{\partial_i'\partial_j}{\partial_l'\partial_l}
  \right),
\quad\!
  P_{\rm L}^{\mu\nu} \!\!\!=\! \eta^{\mu\nu} 
  \!\!\!-\! \frac{{\partial^\mu}'\partial^\nu}{\partial'\!\cdot\partial}
  \!-\!  P_{\rm T}^{\mu\nu}\!\!,\,
\label{ph-m.14}
\end{eqnarray} 
while the transverse and `longitudinal' polarizations are 
\begin{eqnarray} 
 \Pi_{t}(x,x') &=& \partial'\cdot\partial\, 
\Pi^{(1)}_{\rm ren} (x,x')
              + \nabla'\cdot\nabla\, \Pi^{(2)}_{\rm ren}(x,x')\, ,
\nonumber\\
  \Pi_{l}(x,x') &=& 
\partial'\cdot\partial\,\Pi^{(1)}_{\rm ren}(x,x') .
\label{ph-m.15}
\end{eqnarray} 
However, this analogy has its limitations. The absence of 
time-translation invariance in our case makes it non-trivial to
extract local physical quantities such as the photon mass or 
dissipative rate. This is nevertheless possible. 
Below, we perform a perturbative analysis and 
show how one can extract a local photon mass from the 
self-energy~(\ref{ph-m.9})-(\ref{ph-m.12}). 

%\section{Photon mass}

\vskip 0.1in

{\it 4. Photon mass.}
 In order to study the effects of the photon 
self-energy~(\ref{ph-m.9})-(\ref{ph-m.12}) on the photon propagation,
we make use of the Schwinger-Keldysh 
formalism~\cite{SchwingerKeldysh:1961-64,DeWittJordan:1967-86} and write 
the photon field equation of motion as follows: 
\begin{equation}
    \partial_\nu F^{\nu\mu} 
  + \int d^4 x' [^\mu\Pi_{\rm ren}^{{\rm r}\nu}](x,x') A_{\nu} (x')
  = 0 ,
\label{ph-m.17}
\end{equation}
where $[^\mu\Pi_{\rm ren}^{{\rm r}\nu}](x,x') 
  \equiv [^\mu\Pi^{\nu}_{++}](x,x') + [^\mu\Pi^{\nu}_{+-}](x,x')$
%
%\begin{equation}
%         \Pi_{\rm ren}^{{\rm r}\mu\nu}(x,x') 
%  \equiv \Pi^{\mu\nu}_{++}(x,x') + \Pi^{\mu\nu}_{+-}(x,x')
%\label{ph-m.18}
%\end{equation}
%
defines the retarded photon self-energy in terms of the 
Feynman $[^\mu\Pi^{\nu}_{++}]$ and 
Wightman $[^\mu\Pi^{\nu}_{+-}]$ self-energies so that 
the photon propagation is manifestly causal.
The tensors $[^\mu\Pi^{\nu}_{++}]$ and $[^\mu\Pi^{\nu}_{+-}]$ are obtained
from~(\ref{ph-m.9})-(\ref{ph-m.12}) by making the replacements 
$\Delta x^2 \rightarrow  
\Delta x^2_{++}  = - (|\eta-\eta'| - i\epsilon)^2 + 
\mbox{$\vert\vec x-\vec x'\vert^2$}$ and
$\Delta x^2 \rightarrow  
\Delta x^2_{+-} = - (\eta-\eta'   + i\epsilon)^2 + \vert\vec x-\vec x'\vert^2$,
respectively~\cite{ProkopecTornkvistWoodard:2002,DeWittJordan:1967-86}.

Since the vacuum polarization is only known to one loop order we
solve Eq.~(\ref{ph-m.17}) perturbatively, expanding the photon wave
function as
\begin{equation}
 A_\mu = A_\mu^{(0)} + A_\mu^{(1)} + \ldots.
\label{perturbative-A}
\end{equation}
Here $A_\mu^{(1)} = {\cal O}(e^2)$ is the one-loop amplitude,
$A_\mu^{(0)} = \varepsilon_\mu \,e^{ik\cdot x}$ 
(with  $k_0 = \pm |\vec k|$)
is the plane-wave solution to the free Maxwell equation,
and $\varepsilon_\mu$ is the (transverse) photon polarization vector,
which in Lorentz gauge satisfies $\varepsilon_0 = 0$,  $\vec\varepsilon 
\cdot \vec k = 0.$ The one-loop contribution to Eq.~(\ref{ph-m.17}) 
then reads
\begin{equation}
\Big(
  \eta^{\mu\nu} \partial^2 - \partial^{\mu}\partial^{\nu}\!
\Big) A^{(1)}_{\nu} (x)  
  +\! \int \!\! d^{\,4}\! 
x'\, [^\mu\Pi_{\rm ren}^{{\rm r}\nu}](x,x') A^{(0)}_{\nu} (x')
  = 0 .
\label{ph-m.e2}
\end{equation}

We are primarily interested in photons that are subhorizon ($k \gg H$)
at the initial time $\eta_0 = -H^{-1}$, and then become superhorizon at
some later time $\eta$, $k_{\rm phys} = k/a(\eta) \ll H$, as illustrated 
in Fig.~\ref{figure 2}. Upon inserting equations~(\ref{ph-m.9}), 
(\ref{ph-m.11}) and~(\ref{ph-m.12}) into~(\ref{ph-m.e2}), we obtain the 
following approximate equation for the gauge 
field~\cite{ProkopecTornkvistWoodard:2002}
\begin{equation}
\Big(
  \eta^{\mu\nu} \partial^2 - \partial^{\mu}\partial^{\nu}
\Big) A^{(1)}_{\nu}
  - a^2 m_\gamma^2\,\eta^{\mu\nu}\! A^{(0)}_{\nu}
  = 0 ,
\label{ph-m.e2a}
\end{equation}
When evaluated at the leading logarithmic order in $k/H$ and $Ha/k$, the
photon mass-squared is
\begin{equation}
     m_\gamma^2 = \frac{e^2H^2}{2\pi^2}\ln \frac{k}{H}.
\label{ph-m.e2b}
\end{equation}
\begin{figure}[htbp]
\centerline{\hspace{.0in} 
\epsfig{file=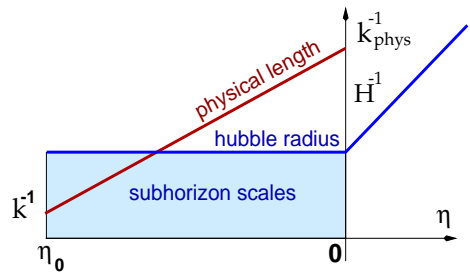, width=2.4in,height=1.4in}
}
\vskip -0.1in
\lbfig{figure 2}
\caption[fig2]{
  \small Evolution of the physical scales in de Sitter inflation.
}
\end{figure}

In what follows we shall discuss the origin and the physical relevance 
of this result. Note first that we have calculated only the leading 
logarithmic contribution to the photon mass. This will be a good 
approximation for modes which satisfy $\ln ({k}/{H}) \gg 1 $.
The mass~(\ref{ph-m.e2b}) corresponds to that of space-like transverse 
excitations, so it is associated with the transverse polarization 
$\Pi_{t}$ in~(\ref{ph-m.15}).

Even though the scalar excitations produced by an inflationary universe
are not thermal, one commonly defines the 'Hawking temperature'
$T_H = H/2\pi$. In terms of this temperature the photon mass-squared 
(\ref{ph-m.e2b}) is $m_\gamma^2 = 2{e^2} T_H^2 \ln (k/H)$. The logarithmic 
enhancement is a consequence of the nonthermal nature of the spectrum of 
charged scalar excitations. 

The mathematics of our photon mass generation mechanism bears an interesting 
resemblance to that of the Schwinger model~\cite{Schwinger:1962} in which
the photon acquires a mass $m_\gamma^2 = e^2/\pi$. In flat, two dimensional,
scalar QED the charged field propagators are logarithmic, which results in the
vacuum polarization failing to vanish on shell. The scalar propagator goes 
like $1/{\Delta x}^2$ in 3+1 dimensional flat space, and the photon stays
massless. In de Sitter background the scalar propagator has a logarithmic 
tail which is responsible for our mass generation effect. The two extra 
spatial dimensions are compensated, in the integration, by two factors of
$1/\eta'$, and the net result is quite similar to Schwinger's.

We now relate the photon mass (\ref{ph-m.e2b}) to the 
Hartree-approximation result 
$(m_\gamma^2)_{\rm Hartree} = 
2e^2 \langle\Omega| \phi^\dagger\phi |\Omega\rangle$
considered in~\cite{DavisDimopoulosProkopecTornkvist:2000}. The Hartree mass
arises from the local contribution to the vacuum polarization~(\ref{ph-m.8}),
represented by the first diagram in Fig.~\ref{figure 1}, and can be
estimated by taking the coincident limit of the propagator~(\ref{ph-m.4}), 
\begin{equation}
 \langle \Omega | \phi^\dagger(x)\phi(x) | \Omega \rangle_{\rm finite}
%   \approx \frac{H^2}{4\pi^2}\, \ln\Big(\frac{a}{a_0}\Big),
   \approx ({H^2}/{4\pi^2})\, 
%\ln({a}/{a_0}),
\ln a~,
\label{coincident}
\end{equation}
where we subtracted the initial (vacuum) contribution at $\eta=\eta_0$
%($a_0\equiv 
[with $a(\eta_0) = 1$].
The Hartree mass exactly agrees with our result (\ref{ph-m.e2b}) if one
makes the reasonable assumption that a mode
with comoving wave number $k$ freezes in at $a(\eta)=k/H$, 
when the mode's physical wavelength redshifts beyond the
causal horizon. 

On the other hand, 
it is premature to claim more than that the field equation
%we have derived, Eq.~
(\ref{ph-m.e2a}) is consistent with 
the presence of a photon mass.
The actual vacuum polarization is nonlocal and, moreover, our perturbative 
result (\ref{ph-m.e2b}) for the mass arises from a 
genuinely nonlocal contribution~\cite{ProkopecTornkvistWoodard:2002}, 
since the local term that gives the Hartree mass is exactly canceled by 
a term from the nonlocal diagram in Fig.~1. 
To prove that superhorizon modes approach the behavior of a massive photon
would require gaining control over higher loop corrections and then solving
the integral-differential equation (\ref{ph-m.17}).
We expect that higher loops induce corrections
at most of the order of $\alpha_e \ln (k/H)$ 
relative to the one-loop perturbative 
result~(\ref{ph-m.e2b}). 
If this conjecture is confirmed, the one-loop
equation~(\ref{ph-m.17}) would suffice to study the photon field dynamics
in inflation, provided that $\alpha_e \ln (k/H)\ll 1$ is satisfied. 
A detailed investigation of this question 
%is 
will be the
subject of future work.     

%Assuming that the entire photon mass is built up during the time when
%the physical photon momenta $k_{\rm phys} = k/a$ correspond to
%subhorizon scales, $k_{\rm phys} \geq H$, 
%we obtain a Hartree mass that is 
%identical to the perturbative result~(\ref{ph-m.e2b}).
%%$(m_\gamma)_{\rm Hartree} = (m_\gamma)_{\rm perturbative}$. 
%Even though this agreement is remarkable, we emphasise that,

%stricly speaking, Eq.~(\ref{ph-m.e2b}) is {\it not} the photon mass.
%To really get the photon mass, a full nonperturbative analysis of 
%(\ref{ph-m.17}) is required. Furthermore, from the perturbative
%analysis we know that the local term in~(\ref{ph-m.8}) 
%responsible for the Hartree mass is {\it cancelled exactly} by a term 
%in the nonlocal diagram in Fig.~\ref{figure 1}, so that 
%the perturbative result is obtained from a genuinely nonlocal
%contribution~\cite{ProkopecTornkvistWoodard:2002}.

 In order to 
%study 
investigate
potentially observable effects of a dynamically
generated photon mass in inflation, it would be necessary to solve the 
%photon dynamical equation 
dynamical equation for the photon
in inflation and match it onto 
solutions for
the radiation and matter
eras. The result may be the generation of cosmological magnetic fields
that seed the galactic dynamo
mechanism~\cite{DavisDimopoulosProkopecTornkvist:2000}. However, in order
to perform a reliable analysis of 
%the 
photon propagation in inflation, 
a more detailed analysis of the photon field equation~(\ref{ph-m.17}) 
is required. We postpone this analysis for a later publication. 

\vskip 0.1in

%\section*{Acknowledgements}

{\it 5. Acknowledgements.}
We wish to thank Anne Davis and Konstantinos Dimopoulos for work on related
issues, and Misha Shaposhnikov for pointing out possible limitations of 
the Hartree approximation. This work was partially supported by 
DOE contract DE-FG02-97ER41029 and by the IFT
%Institute for Fundamental Theory 
of the University of Florida.

%
%%%%%%%%%%%%%%%%%%%%%%%%%%%%%%%%%%%%%%%%%%%%%%%%%%%%%%%%%%%%%%%%%%%%%%%%%%%%%%%
%    REFERENCES
%%%%%%%%%%%%%%%%%%%%%%%%%%%%%%%%%%%%%%%%%%%%%%%%%%%%%%%%%%%%%%%%%%%%%%%%%%%%%%%
%

\end{document}